\documentclass[lettersize,journal]{IEEEtran}
\IEEEoverridecommandlockouts
\usepackage{amsmath,amsfonts}
\usepackage{booktabs}
\usepackage{algorithmic}
\usepackage{algorithm}
\usepackage{array}
\usepackage[caption=false,font=normalsize,textfont=sf]{subfig}
\usepackage{textcomp}
\usepackage{stfloats}
\usepackage{url}
\usepackage{verbatim}
\usepackage{graphicx}
\usepackage{cite}
\usepackage{multicol}
\usepackage{amssymb}
\usepackage{lineno}
\usepackage{makecell}
\usepackage{indentfirst}
\usepackage{bm}
\usepackage{color}
\usepackage{cite}
\usepackage{epstopdf}
\usepackage{amssymb}
\usepackage{xcolor}
\usepackage{tikz}
\usepackage{mathtools}
\usepackage{pifont}
\usepackage{marvosym}
\usepackage{newtxmath}
\usepackage[colorlinks,
linkcolor=blue,
anchorcolor=blue,
citecolor=blue]{hyperref}

\newcommand{\diag}{\mathrm{diag}}

\newcommand{\tr}{\mathrm{tr}}
\newcommand{\HH}{\mathrm{H}}

\newtheorem{proposition}{\textbf{Proposition}}

\newtheorem{lemma}{\textbf{Lemma}}

\begin{document}
	
	\title{\huge Sensing Mutual Information for   Communication Signal with Deterministic Pilots and Random Data Payloads}

\author{Lei Xie,~\IEEEmembership{Member,~IEEE}, Hengtao He,~\IEEEmembership{Member,~IEEE}, Jun Tong,~\IEEEmembership{Member,~IEEE},\\
Fan Liu,~\IEEEmembership{Senior Member,~IEEE}, and Shenghui Song,~\IEEEmembership{Senior Member,~IEEE}
		\thanks{L. Xie is with School of Cyber Science and Engineering, Southeast University, Nanjing 210096, China. H. He and F. Liu are with School of Information Science and Engineering, Southeast University, Nanjing 210096, China. J. Tong is with the School of Electrical, Computer, and Telecommunications Engineering, University of Wollongong, Wollongong, NSW 2522, Australia. S. Song is with the Division of Integrative Systems and Design and the Department of Electronic and Computer Engineering, the Hong Kong University of Science and Technology, Hong Kong.}
	}
	
	\maketitle
	
	\begin{abstract}
	The recent emergence of the integrated sensing and communication (ISAC) framework has sparked significant interest in quantifying the sensing capabilities inherent in communication signals. However, existing literature has mainly focused on scenarios involving either purely random or purely deterministic waveforms. This overlooks a critical reality: operational communication standards invariably utilize a hybrid structure comprising both deterministic pilots for channel estimation and random payloads for data transmission. To bridge this gap, this paper investigates the sensing mutual information (SMI) and precoding design specifically for ISAC systems employing communication signals with both pilots and data payloads. First, by utilizing random matrix theory (RMT), we derive a tractable closed-form expression for the SMI that accurately accounts for the statistical properties of the hybrid signal. Building upon this theoretical foundation, we formulate a precoding optimization problem to maximize SMI with constraints on the transmit power and communication rate, which is solved via an efficient alternating direction method of multipliers framework. Simulation results validate the accuracy of the theoretical results and demonstrate the superiority of the proposed precoding design over conventional benchmarks. 
	\end{abstract}
	
	\begin{IEEEkeywords}
		Integrated sensing and communication, precoder design, random signal, sensing mutual information.
	\end{IEEEkeywords}
	
	\section{Introduction}
As the low-altitude economy accelerates the deployment of mission-critical applications, such as autonomous UAV logistics, urban air mobility, and large drone swarms, the demand for high-fidelity environmental sensing has transformed sensing into a fundamental functionality for future integrated communication and sensing (ISAC) networks  \cite{liu2022integrated,xie2023collaborative,cheng2025networked}.  
	However, there are still substantial challenges that need to be tackled before the potential of ISAC can be fully unleashed. For example, radar and communication systems have diverged in their signal requirements: radar relies on deterministic waveforms for stability, while communication demands randomness for information transfer \cite{xiong2023fundamental,li2025rethinking}. Although ISAC studies have examined sensing with random signals \cite{lu2023random,xie2023sensing}, they often overlook the hybrid nature of practical communication signals, which invariably mix deterministic pilots with random data payloads.

	To address this issue, a few recent studies have extended their analysis to such hybrid signaling schemes \cite{xu2025exploiting,xie2025bistatic,xie2025adaptive}. In particular, the authors of  \cite{xu2025exploiting} evaluated estimation performance, while \cite{xie2025bistatic} focused on target detection, proposing a generalized likelihood ratio test (GLRT) detector to validate that both deterministic and random components improve reliability. Furthermore, \cite{xie2025adaptive} addressed clutter mitigation through adaptive filtering. Despite these advances, a disconnect remains: the metrics used in these studies, e.g., the ergodic linear minimum mean square error (ELMMSE), the detection probability, and the signal-to-clutter-and-noise ratio, are often incompatible with conventional communication metrics. Consequently, establishing a unified performance metric that captures the essence of both sensing and communication within a hybrid signal framework remains an open challenge.
		
	This paper pursues two primary objectives: 1) to evaluate the sensing performance for communication signals with both deterministic pilots and random data payloads by sensing mutual information (SMI), which is compatible with the communication performance metric mutual information; and 2) to optimize the ISAC performance via precoding design. To achieve the first objective, we derive a closed-form expression for the SMI between the random target response and the received signals, accounting for the hybrid nature of the transmit waveform. Building upon this theoretical analysis, we maximize the SMI with constraints on the transmit power and communication rate and optimize the transmit signal by utilizing the alternating direction method of multipliers (ADMM) framework. Simulation results validate the accuracy of the derived expression and demonstrate that the proposed method outperforms the existing benchmarks.

	The remainder of the paper is organized as follows.  
	Sec. II presents the system model. Sec. III provides a closed-form expression for the SMI. 
	Sec. IV introduces an ADMM-based approach to maximize the SMI by optimizing the precoding. 
	Simulation results are given in Sec. V to validate the accuracy of the theoretical analysis and the effectiveness of the proposed method.	
	Finally, Sec. VI concludes the paper.

\begin{figure}[!t]
\centering\includegraphics[width=2.6in]{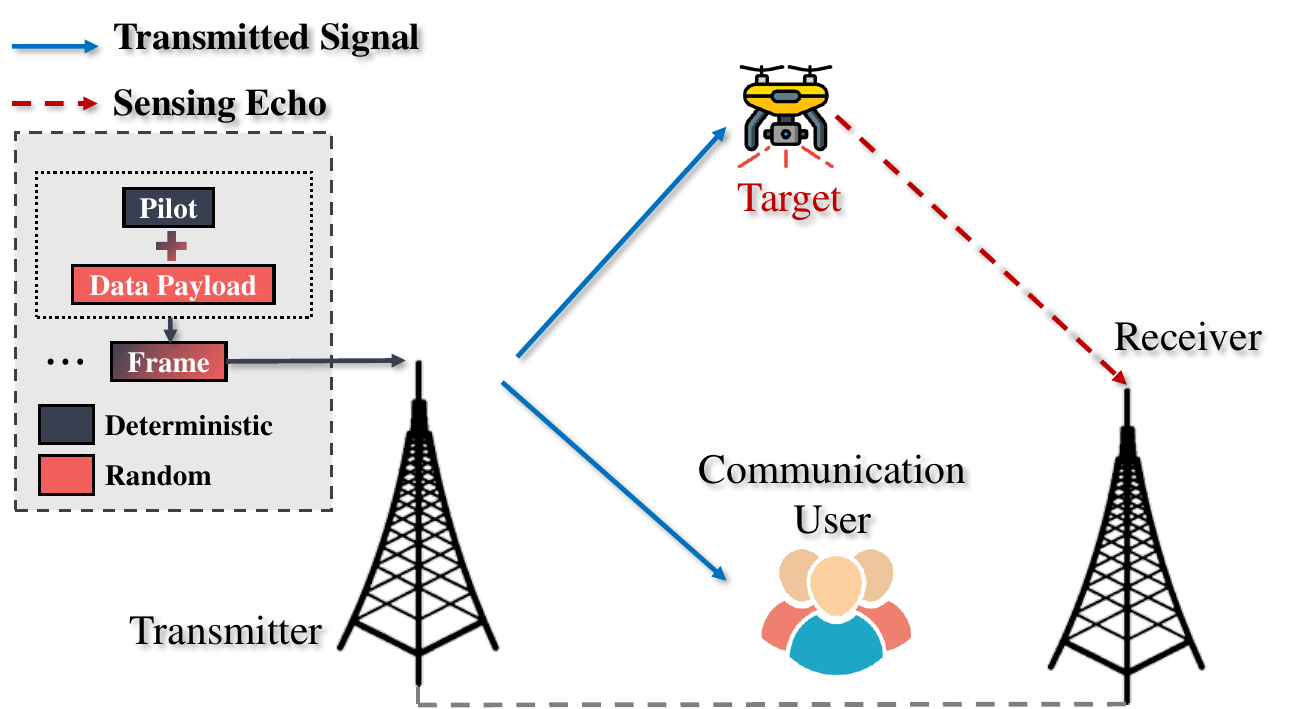}
	\caption{Illustration of the ISAC system.}
	\label{SYsill}
\end{figure}

	\section{System Model}

    Consider a bi-static sensing system, which is composed of a sensing transmitter equipped with $N_{t}$ antennas and a sensing receiver with $N_{r}$ antennas. Assume that the target sensing is performed within a coherent processing interval (CPI) consisting of $L$ time slots, as illustrated in Fig. \ref{SYsill}. 	

	In this work, we adopt the superimposed pilot (SIP) transmission scheme \cite{xie2024superimposed,xie2025adaptive} to improve the spectrum efficiency. In particular, the transmitted signal comprises both pilot and data payload, i.e., 
	\begin{equation}
		\mathbf{X} = \mathbf{S}_p + \mathbf{F}\mathbf{S}_d \in \mathbb{C}^{N_{t}\times L},
	\end{equation}
	where $\mathbf{F}\in \mathbb{C}^{N_{t}\times N_{t}}$
	denotes the precoding matrix, $ \mathbf{S}_p, \mathbf{S}_d\in \mathbb{C}^{N_{t}\times L}$ represents the deterministic pilot and the random data payload, respectively. In particular, the entries of $\mathbf{S}_d$ are assumed to be independent and identically distributed (i.i.d.) complex Gaussian random variables with zero mean and unit variance. To minimize the error of channel estimation, the deterministic pilots of different transmit antennas are designed to be omnidirectional and mutually orthogonal, i.e., \cite{xu2025exploiting}
	 \begin{equation}
	 	\mathbf{S}_p \mathbf{S}_p^\HH = \mathbf{I}_L,
	 \end{equation}
	 where $(\cdot)^\HH$ denotes the Hermitian transpose.
	  
	  Consider a single target and then the  target response matrix is given by
	\begin{equation}
		\mathbf{H} = \alpha \mathbf{a}_r \mathbf{a}_t^\HH  \in \mathbb{C}^{N_{r}\times N_{t}},
	\end{equation}
	where $\alpha$ is the complex path gain, accounting for the radar cross section (RCS) and path loss \cite{xie2020recursive}. The vectors $\mathbf{a}_t$ and $\mathbf{a}_r$ denote the transmit and receive steering vectors, respectively. The received signal at the sensing receiver is given by 
	\begin{equation}
		\begin{split}
			\mathbf{Y}_s =   \mathbf{H} \mathbf{X}+ \mathbf{N}_s\in \mathbb{C}^{N_{r}\times L},
		\end{split}
	\end{equation}
	where $\mathbf{N}_s \in \mathbb{C}^{N_{r}\times L}$ represents additive white Gaussian noise (AWGN) with noise power $\sigma^2$. 
	
	By vectorizing the conjugate transpose of the received signal matrix $\mathbf{Y}_s$, we obtain the equivalent system model, i.e.,
	\begin{equation}
		\begin{split}
			\mathbf{y}_s \triangleq \text{vec} \left(\mathbf{Y}_s^\HH\right)= \left(\mathbf{I}_{N_{r}} \otimes \mathbf{X}\right)^\HH \mathbf{h} + \mathbf{n} \in \mathbb{C}^{N\times 1},
		\end{split}
	\end{equation}
	where $N = N_{t}N_{r}$, $\mathbf{h}= \text{vec}(\mathbf{H}^\HH)$, and the noise vector satisfies $\mathbf{n}_s= \text{vec}(\mathbf{N}_s^\HH)\sim\mathcal{CN}(\mathbf{0},\sigma^2\mathbf{I}_N)$. 	
	The correlation matrix of $\mathbf{h}$ is defined as 
\begin{equation}
	\begin{split}
\mathbf{R}\triangleq \mathbb{E}\left(\mathbf{h}\mathbf{h}^\HH\right) = |\alpha|^2\mathbf{R}_{r} \otimes \mathbf{R}_{T},
	\end{split}
\end{equation} 
where $\mathbf{R}_{r} = \mathbf{a}_r\mathbf{a}_r^\HH$ and $\mathbf{R}_{T}= \mathbf{a}_t\mathbf{a}_t^\HH$ denote the spatial correlation matrices at the receiver and the transmitter, respectively. 
	 
	 The communication signal is defined by $\mathbf{Y}_c = \mathbf{G}\mathbf{X} + \mathbf{N}_c$, 
	 where $\mathbf{G}$ represents the communication channel and $\mathbf{N}_c$ denotes the AWGN with noise power $\sigma^2$.
	 The communication performance is measured by the channel capacity, i.e.,
\begin{equation}	
	I_c=\log \left\vert\mathbf{I}+\frac{1}{L\sigma^2}\mathbf{G} \mathbf{F}\mathbf{S}_d\mathbf{S}_d^\HH\mathbf{F}^\HH \mathbf{G}^\HH\right\vert = \log \left\vert\mathbf{I}+\frac{1}{\sigma^2}\mathbf{G} \mathbf{\Theta} \mathbf{G}^\HH\right\vert,
\end{equation}
	where $\mathbf{\Theta} = \mathbf{F}\mathbf{F}^\HH$ denotes the correlation matrix of the precoder.

	\section{SMI for communication signals with both deterministic pilots and random data payloads}

To detect the target, the sensing receiver estimates the target response vector $\mathbf{h}$ from the received signal $\mathbf{y}$. Prior works that assume deterministic sensing waveforms optimize $\mathbf{S}$ to maximize the SMI between the received signal and the target response \cite{xie2023sensing}. However, such approaches cannot be directly extended to ISAC systems, where the transmit waveform contains random components.

Given that $\mathbf{h}$ follows a Gaussian distribution \cite{1300860}, the SMI for communication signals is defined as \cite{xie2023sensing}
\begin{equation}\label{SEMIdef}
	\begin{split}
		&I_s\triangleq \mathbb{E}_{\mathbf{X}}\log \left\vert\mathbf{I}+ \frac{|\alpha|^2}{L\sigma^{2}}\left(\mathbf{R}_{r} \otimes \mathbf{R}_{t}^{\frac{1}{2}}\mathbf{X} \mathbf{X}^\HH\mathbf{R}_{t}^{\frac{1}{2}}\right) \right\vert,
	\end{split}
\end{equation}
where the expectation is taken over the random signals $\mathbf{X}$. 
While the preliminary results in \cite{xie2023sensing} derived the SMI under the assumption of purely random signals where $\mathbf{X}$ is a Gaussian matrix, a closed-form expression for the SMI for the hybrid signal with both deterministic and random components remains an intractable challenge.
This is because the presence of deterministic components alters the statistical properties of the matrix $\mathbf{X}\mathbf{X}^\HH$ from a central Wishart distribution to a non-central one, which significantly complicates the statistical analytical framework.

	To address this issue, we establish a closed-form expression for the SMI by examining the asymptotic behavior of  $I\left(\mathbf{h};\mathbf{y}\right)$. Specifically, we present an approximation for the SMI in the large-$L$ regime, utilizing a first-order approximation of the MI. The main result is summarized in the following proposition.
	
	\begin{proposition}\label{TheoRD}
		Let $(\delta,\tilde{\delta})$ satisfy the following system of fixed-point equations:

				\begin{equation}\label{fixeq1}
			\left\{\begin{aligned}
				\delta &= \frac{1}{L} \frac{d}{\rho^{-1}(1+\tilde{\delta}d)+ \frac{\Vert\mathbf{a}_t\Vert^2}{L(1+\delta)}}, \\
				\tilde{\delta} &= \frac{1}{L} \frac{1}{\rho^{-1}(1+\delta)+\frac{\Vert\mathbf{a}_t\Vert^2}{L(1+\tilde{\delta}d)} }+\frac{L-1}{L}\rho(1+\delta)^{-1},
			\end{aligned}\right.
		\end{equation}
		where $\rho = |\alpha|^2\sigma^{-2} \tr(\mathbf{R}_{r})$ and $d = \mathbf{a}_t^\HH \mathbf{\Theta} \mathbf{a}_t$.
		In the asymptotic region where $L \to \infty$, the SMI is given by
		\begin{equation}\label{SEMI000}
			I_s = \bar{I}_s + \mathcal{O}\left({L}^{-1}\right),
		\end{equation}
		where the deterministic approximation $\bar{I}_s$ is defined as
		\begin{equation}\label{SEMI0}
			\begin{split} 
				\bar{I}_s \triangleq \log\left\vert 1+\tilde{\delta}d+ \frac{\rho \Vert\mathbf{a}_t\Vert^2}{L(1+\delta)}  \right\vert + L\log(1+\delta) - \rho^{-1}L\delta \tilde{\delta}.
			\end{split}
		\end{equation}
	\end{proposition}
	
	\textbf{\emph{Proof:}} 	This proof necessitates an analysis of the asymptotic behavior of the random variable $I_s$. Our argument adapts the framework from random matrix theory, specifically extending the results presented in \cite[Theorem 1]{5429113}. For completeness and ease of reference, the relevant result is summarized in the following lemma.
	
	\begin{lemma}[{\cite[Theorem 1]{5429113}}]
		\label{lem:det_eq}
		Consider a matrix $\mathbf{\Sigma} = \mathbf{B}+\mathbf{Y} \in \mathbb{C}^{r\times t}$. Assume the random component is modeled as $\mathbf{Y} = \frac{1}{\sqrt{t}} \mathbf{D}^{\frac{1}{2}}\mathbf{X}\widetilde{\mathbf{D}}^{\frac{1}{2}}$, where $\mathbf{D} = \diag(d_i)_{i=1}^r$ and $\widetilde{\mathbf{D}} = \diag(\tilde{d}_j)_{j=1}^t$ are deterministic diagonal matrices, and $\mathbf{X} \in \mathbb{C}^{r \times t}$ contains i.i.d. entries with zero mean and unit variance. Let $(\delta, \tilde{\delta})$ be the unique positive solution to the following system of fixed-point equations:
		\begin{equation}
			\begin{cases}
				\delta = \frac{1}{t} \tr\left(\mathbf{D}\mathbf{\Gamma}\right), \\
				\tilde{\delta} = \frac{1}{t}\tr\left(\widetilde{\mathbf{D}}\widetilde{\mathbf{\Gamma}}\right),
			\end{cases}
		\end{equation}
		where the auxiliary matrices are defined as
		\begin{equation}
			\begin{cases}
				\mathbf{\Gamma} = \left[\sigma^2 \mathbf{\Phi}^{-1}+\mathbf{B}\widetilde{\mathbf{\Phi}}\mathbf{B}^\HH\right]^{-1}, \\ 
				\widetilde{\mathbf{\Gamma}} = \left[\sigma^2 \widetilde{\mathbf{\Phi}}^{-1}+\mathbf{B}^\HH\mathbf{\Phi}\mathbf{B}\right]^{-1}, \\
				\mathbf{\Phi} = \left(\mathbf{I}_r + \tilde{\delta}\mathbf{D}\right)^{-1}, \\
				\widetilde{\mathbf{\Phi}} = \left(\mathbf{I}_t + \delta\widetilde{\mathbf{D}}\right)^{-1}.
			\end{cases}
		\end{equation}
		Under these conditions, the following deterministic approximation holds for the ergodic capacity
		\begin{equation}
			\mathbb{E}_{\mathbf{X}}\left[\log\det\left(\mathbf{I}_r + \sigma^{-2}\mathbf{\Sigma}\mathbf{\Sigma}^\HH\right)\right] = \overline{I} + \mathcal{O}\left(\frac{1}{t}\right),
		\end{equation}
		where the asymptotic value $\overline{I}$ is given by
		\begin{equation}
			\overline{I} = - \log \det\left(\sigma^{2}\mathbf{\Gamma}\right) + \log \det\left(\mathbf{I}_t + \delta\widetilde{\mathbf{D}}\right) - \sigma^2 t \delta \tilde{\delta}. 
		\end{equation}
	\end{lemma}
	
	The proper evaluation of $I_s$ requires recasting it into an extended model that aligns with the structure of \textbf{Lemma \ref{lem:det_eq}}. Based on the definitions of the spatial correlation matrices $\mathbf{R}_{t}$ and $\mathbf{R}_{r}$, we can obtain
	\begin{equation}
		I_s = \log \left( 1+\rho\mathbf{z}\mathbf{z}^\HH\right),
	\end{equation}
	where we introduce the effective channel vector $\mathbf{z}$ as
	\begin{equation}
		\mathbf{z} = \frac{1}{\sqrt{L}} \mathbf{a}_t^\HH \mathbf{X} \in \mathbb{C}^{1\times L}.
	\end{equation}
	To isolate the deterministic and random components, we define the deterministic mean vector as $\mathbf{b} = \frac{1}{\sqrt{L}} \mathbf{a}_t^\HH \mathbf{S}_p$. The $l$-th entry of $\mathbf{z}$ follows a Gaussian distribution with mean $[\mathbf{b}]_l$ and variance given by
	\begin{equation}
		\sigma_{z,l}^2 = \frac{1}{L} \mathbf{a}_t^\HH \mathbf{F}\mathbf{F}^\HH \mathbf{a}_t = \frac{d}{L}.
	\end{equation}
	Consequently, we can represent $\mathbf{z}$ in a form compatible with the random matrix model
	\begin{equation}
		\mathbf{z} \sim \mathbf{b} + \frac{1}{\sqrt{L}} d^{\frac{1}{2}} \mathbf{x} \mathbf{I}_L^{\frac{1}{2}},
	\end{equation} 
	where $\mathbf{x} \in \mathbb{C}^{1\times L}$ is a vector of i.i.d. entries with zero mean and unit variance. This representation allows us to invoke \textbf{Lemma \ref{lem:det_eq}} by mapping the parameters as follows: set dimensions $r=1$ and $t=L$, and assign matrices $\mathbf{B} = \mathbf{b}$, $\mathbf{D} = d$, and $\widetilde{\mathbf{D}} = \mathbf{I}_L$. It follows that		
	\begin{equation}
		\begin{cases}
			\delta = \frac{1}{L} d\Gamma, \\
			\tilde{\delta} = \frac{1}{L}\tr\left(\widetilde{\mathbf{\Gamma}}\right),
		\end{cases}
	\end{equation}
	where the auxiliary matrices are defined as
	\begin{equation}\label{4fpe}
		\begin{cases}
			\Gamma = \left[\rho^{-1} \phi^{-1}+\mathbf{b}\widetilde{\mathbf{\Phi}}\mathbf{b}^\HH\right]^{-1}, \\ 
			\widetilde{\mathbf{\Gamma}} = \left[\rho^{-1} \widetilde{\mathbf{\Phi}}^{-1}+\phi\mathbf{b}^\HH\mathbf{b}\right]^{-1}, \\
			\phi = \left(1 + \tilde{\delta}d\right)^{-1}, \\
			\widetilde{\mathbf{\Phi}} = \left(1 + \delta\right)^{-1} \mathbf{I}_L.
		\end{cases}
	\end{equation}
	Then, \eqref{4fpe} indicates that
\begin{equation}\label{GammaDef}
\begin{cases}
		\Gamma =  
		\left[\rho^{-1}(1+\tilde{\delta}d)+ \frac{\Vert\mathbf{a}_t\Vert^2}{L(1+\delta)}\right]^{-1},\\
		\widetilde{\mathbf{\Gamma}} =  \left[\rho^{-1}(1+\delta)\mathbf{I} + \frac{1}{L(1+\tilde{\delta}d)}\mathbf{S}_p^\HH  \mathbf{a}_t \mathbf{a}_t^\HH\mathbf{S}_p\right]^{-1}.
\end{cases}
\end{equation}
	The asymptotic deterministic equivalent for $I_s$ is given by 
	\begin{equation}
		I_s \overset{a.s.}{\to} -\log \left(\rho^{-1}\Gamma\right) + \log \det\left(\mathbf{I}_L+\delta\widetilde{\mathbf{D}}\right) - \sigma^2  L \delta \tilde{\delta} = \bar{I}_s,
	\end{equation}
	where the pair $(\delta, \tilde{\delta})$ is obtained as the unique solution to the fixed-point equation defined in \eqref{fixeq1}. \hfill $\blacksquare$

	\emph{Proposition \ref{TheoRD}} establishes a closed-form expression for the SMI between the target response vector and the received signals. As shown in the subsequent sections, although \eqref{SEMI0} is derived in the asymptotic region, it provides an accurate approximation of the SMI even for finite $L$.

	\section{SMI-Oriented Precoding Design}
		Next, we optimize the ISAC precoder to maximize the SMI. The optimization problem is formulated as follows:
		\begin{equation}\label{P0}
			\begin{array}{ccl}
				\mathcal{P}_0:&\max\limits_{\mathbf{\Theta}} &I_s\left(\mathbf{\Theta}\right)\\
				& s.t. & \tr\left(\mathbf{\Theta}\right) \leq P, \; I_c\left(\mathbf{\Theta}\right) \geq R_0,
			\end{array}
		\end{equation}
		where the transmit power is constrained by a maximum value $P$, and the communication performance is constrained to be larger than a given level $R_0$. Since the variable $(\delta,\tilde{\delta})$ is derived implicitly by solving the fixed-point equation in \eqref{fixeq1}, it is hard to solve $\mathcal{P}_0$ directly. To this end, we exploit ADMM by introducing an auxiliary variable $\mathbf{\Omega}$, such that $\mathcal{P}_0$ can be recast as
		\begin{equation}\label{P1}
		\begin{array}{ccl}
			\mathcal{P}_1:&\max\limits_{\mathbf{\Theta}} &I_s\left(\mathbf{\Theta}\right)\\
			& s.t. & \tr\left(\mathbf{\Theta}\right) \leq P, \; I_c\left(\mathbf{\Omega}\right) \geq R_0, \; \mathbf{\Theta} = \mathbf{\Omega}.
		\end{array}
	\end{equation}
		The problem leads to the augmented Lagrangian function 
		\begin{equation}
			L( \mathbf{\Theta},\mathbf{\Omega},\mathbf{U})= - I_s (\mathbf{\Theta}) + \frac{\gamma}{2}\left\Vert\mathbf{\Theta}-\mathbf{\Omega}+\mathbf{U}\right\Vert^2,
		\end{equation}
		where $\gamma$ is a penalty parameter. Then, at the $m$-th iteration, the variables are updated by
		\begin{subequations}\label{ADMMiter}
		\begin{align}
			&\mathbf{\Theta}^{(m+1)} = \mathop{\arg\min}_{\tr\left(\mathbf{\Theta}\right) \leq P} L\left( \mathbf{\Theta},\mathbf{\Omega}^{(m)},\mathbf{U}^{(m)}\right),\label{subpro_1}\\
			&\mathbf{\Omega}^{(m+1)} = \mathop{\arg\min}_{ I_c\left(\mathbf{\Omega}\right) \geq R_0} L\left( \mathbf{\Theta}^{(m+1)},\mathbf{\Omega},\mathbf{U}^{(m)}\right),\label{subpro_2}\\
			&\mathbf{U}^{(m+1)} = \mathbf{U}^{(m)} + \mathbf{\Theta}^{(m+1)}-\mathbf{\Omega}^{(m+1)},\label{subpro_3}
		\end{align}
		\end{subequations}		
		where $\mathbf{\Theta}^{(m)}$, $\mathbf{\Omega}^{(m)}$, and  $\mathbf{U}^{(m)}$ denote the variables at the $m$-th iteration.
		
		\subsubsection{Updata $\mathbf{\Theta}^{(m+1)}$}
		Given \eqref{subpro_1} is convex, we propose to solve it by a gradient projection (GP)-based method \cite{xie2023sensing}. 
	 To facilitate this, we first present the derivatives of $(\delta,\tilde{\delta})$ with respect to $\mathbf{\Theta}$.
		
		\begin{proposition}\label{lemmaderiveRs}
			Define
\begin{align}
	&a_{11} = 1 +  \frac{d ||\mathbf{a}_t||^2}{L^3(1+\delta)^2{\left(\rho^{-1}(1+\tilde{\delta}d)+ \frac{\Vert\mathbf{a}_t\Vert^2}{L(1+\delta)}\right)^2}},\\
	&a_{12} =   \frac{\rho^{-1} d^2}{\left(\rho^{-1}(1+\tilde{\delta}d)+ \frac{\Vert\mathbf{a}_t\Vert^2}{L(1+\delta)}\right)^2},\\
	&a_{21} = \frac{\rho^{-1}}{L\left(\rho^{-1}(1+\delta)+\frac{\Vert\mathbf{a}_t\Vert^2}{L(1+\tilde{\delta}d)}\right)^2} + \frac{L-1}{L(1+\delta)^2},\\
	&a_{22} = 1 - \frac{||\mathbf{a}_t||^2d }{L^3(1+\tilde{\delta}d)^2\left(\rho^{-1}(1+\delta)+\frac{\Vert\mathbf{a}_t\Vert^2}{L(1+\tilde{\delta}d)}\right)^2},\\
	&b_1 = \frac{1}{L\left(\rho^{-1}(1+\tilde{\delta}d)+ \frac{\Vert\mathbf{a}_t\Vert^2}{L(1+\delta)}\right)}, \\
	&b_2 = \frac{||\mathbf{a}_t||^2\tilde{\delta} }{L^3(1+\tilde{\delta}d)^2\left(\rho^{-1}(1+\delta)+\frac{\Vert\mathbf{a}_t\Vert^2}{L(1+\tilde{\delta}d)}\right)^2}.
\end{align}
			The derivatives of $(\delta,\tilde{\delta})$ are given by
\begin{equation}\label{Pro2}
	\begin{split}
		\left\{
		\begin{aligned}
		\mathbf{\Delta} = \nabla_{\mathbf{\Theta}} \delta = \frac{\partial \delta}{\partial \mathbf{\Theta}}  = c_1 \mathbf{a}_t \mathbf{a}_t^\HH,\\
		\widetilde{\mathbf{\Delta}} = \nabla_{\mathbf{\Theta}} \tilde{\delta} = \frac{\partial \tilde{\delta}}{\partial \mathbf{\Theta}}  = c_2 \mathbf{a}_t \mathbf{a}_t^\HH,
		\end{aligned} \right.
	\end{split}
\end{equation}
			where
			\begin{equation}
				\begin{split}
					\left[\begin{matrix}
						c_1\\
						c_2
					\end{matrix}\right] = \left[\begin{matrix}
						a_{11} & a_{12}\\
						a_{21} & a_{22}
					\end{matrix}\right]^{-1} \left[\begin{matrix}
						b_1\\
						b_2
					\end{matrix}\right].
				\end{split}
			\end{equation}
		\end{proposition}
		
		\textbf{\emph{Proof:}} 
	Before proceeding, the gradient of $d$ with respect to  $\mathbf{\Theta}$ is given by
\begin{equation}
	\begin{split}
		\mathbf{D} = \nabla_{\mathbf{\Theta}}d = \frac{\partial d}{\partial \mathbf{\Theta}}=  \mathbf{a}_t \mathbf{a}_t^\HH.
	\end{split}
\end{equation}

From \eqref{fixeq1}, we have
\begin{equation}\label{delta_2}
	\begin{split}	
		\mathbf{\Delta} =  \frac{1}{L}\left\{\Gamma \mathbf{D}  -d\Gamma^2 \left[\rho^{-1}\left(d\widetilde{\mathbf{\Delta}}+\tilde{\delta}\mathbf{D}\right) - \frac{||\mathbf{a}_t||^2}{L^2(1+\delta)^2}\mathbf{\Delta}\right] \right\},\\
	\end{split}
\end{equation}
\begin{equation}\label{tildedelta_2}
	\begin{split}	
		\widetilde{\mathbf{\Delta}} = -\frac{\left[\rho^{-1}\mathbf{\Delta} - \frac{||\mathbf{a}_t||^2}{L^2(1+\tilde{\delta}d)^2} \left(d\widetilde{\mathbf{\Delta}}+\tilde{\delta}\mathbf{D}\right) \right]}{L\left(\rho^{-1}(1+\delta)+\frac{\Vert\mathbf{a}_t\Vert^2}{L(1+\tilde{\delta}d)}\right)^2} - \frac{(L-1)}{L(1+\delta)^2}\mathbf{\Delta},
	\end{split}
\end{equation}
where $\Gamma $ is defined by \eqref{GammaDef}. 

By rearranging \eqref{delta_2} and \eqref{tildedelta_2}, the following system of linear equations can be obtained, i.e.,
\begin{equation}\label{eq_app}
	\left\{
	\begin{aligned}
		a_{11} \mathbf{\Delta} + a_{12} \widetilde{\mathbf{\Delta}} = b_1 \mathbf{D},\\
		a_{21} \mathbf{\Delta} + a_{22} \widetilde{\mathbf{\Delta}} = b_2 \mathbf{D}.
	\end{aligned}\right.
\end{equation}
Solving the equations leads directly to the result in \eqref{Pro2}. \hfill $\blacksquare$
		
		Then, from \eqref{SEMI0}, we have
		\begin{equation}
			\begin{split}
				\frac{\partial \bar{I}_s}{\partial \mathbf{\Theta}} = & \frac{\left[\rho^{-1}\left(d\widetilde{\mathbf{\Delta}}+\tilde{\delta}\mathbf{a}_t \mathbf{a}_t^\HH\right) - \frac{||\mathbf{a}_t||^2}{L^2(1+\delta)^2}\mathbf{\Delta}\right]}{\rho^{-1}(1+\tilde{\delta}d)+ \frac{\Vert\mathbf{a}_t\Vert^2}{L(1+\delta)}}  +  \frac{L}{(1+\delta)}\mathbf{\Delta}\\
				& - \rho^{-1} L \left(\delta \widetilde{\mathbf{\Delta}} + \tilde{\delta} \mathbf{\Delta} \right).
			\end{split}
		\end{equation}
		Subsequently, at the $m$-th iteration, the variable is updated to $\widehat{\mathbf{\Theta}}_{m+1}$ using:
		\begin{equation}\label{hatSigmam1}
			\widehat{\mathbf{\Theta}}_{m+1} = \mathbf{\Theta}_{m+1} - \beta_m \nabla_{\mathbf{\Theta}} I_s(\mathbf{\Theta}_m).
		\end{equation}
		
		A retraction step is required to map the updated points from the tangent space back onto the constraint manifold. The retraction is defined as:
		\begin{equation}\label{retractionF}
			\begin{split}
				\mathbf{\Theta}_{m+1}=\left\{
				\begin{matrix}
					\widehat{\mathbf{\Theta}}_{m+1}, & \mathrm{if} \; \tr(\widehat{\mathbf{\Theta}}_{m+1}) \leq P,\\
					\frac{P}{\tr(\widehat{\mathbf{\Theta}}_{m+1})}\widehat{\mathbf{\Theta}}_{m+1}, & \mathrm{otherwise}.
				\end{matrix}\right.
			\end{split}
		\end{equation}

		\subsubsection{Update $\mathbf{\Omega}^{(m+1)}$} The subproblem is a typical semi-definite programming (SDP) problem because the objective function and all constraints are reformulated as a linear or quadratic form. This problem can be easily solved by the CVX toolbox \cite{cvx}.

		The convergence of the iteration in \eqref{ADMMiter} is guaranteed by \cite{neal2011distributed}. 
		By alternatively updating $\mathbf{\Theta}$, $\mathbf{\Omega}$, and $\mathbf{U}$, the variable $\mathbf{\Theta}$ will converge to a stationary point as $m$ increases. Subsequently, the precoding matrix $\mathbf{F}$ can be derived by performing Singular Value Decomposition (SVD) on $\mathbf{\Theta}$.

		\section{Simulation Result}
		In this section, numerical results are presented to validate the accuracy of the theoretical analysis and evaluate the effectiveness of the proposed manifold-based algorithms. 
		We consider a millimeter-wave (mmWave) system operating at a carrier frequency of $28$ GHz. The angle of departure (AOD) and angle of arrival (AOA) of the target is set as $10^\circ$ and $20^\circ$. 
		The number of transmit and receive antennas is set to $N_{t} = 4$ and $N_{r} = 6$, respectively.
		For the proposed optimization algorithm, the maximum number of iterations is set to $200$, with a termination tolerance of $10^{-10}$ for the gradient norm between successive iterations. 
		Regarding the power budget, the transmission power is $P = 30$ dBm and the noise power is $\sigma^2 = -90$ dBm. The step size of GP method is set as $\beta_m = 0.01$, the penalty parameter is $\gamma = 10$,

		\begin{figure}[!t]
			\centering
			\includegraphics[width=3.5in]{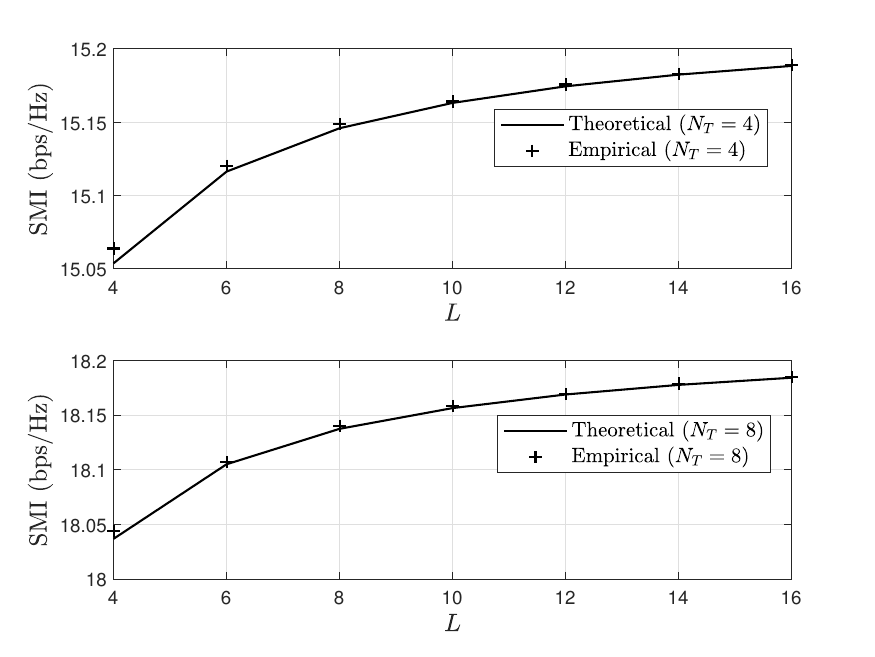}
			\caption{SMI versus the number of time slots $L$. 
			}
			\label{fig_fitness}
		\end{figure}

		Fig.~\ref{fig_fitness} corroborates the accuracy of the closed-form SMI expression in \emph{Proposition~\ref{TheoRD}} for a representative scenario with $K=7$ targets. The curves labeled ``Empirical'' depict the SMI estimated from $N_{\mathrm{mc}}=10{,}000$ Monte-Carlo trials, whereas the curves labeled ``Theoretical'' are obtained from the analytical approximation in~\eqref{SEMI0}. It is observed that the analytical approximation matches the simulation results closely, even in the moderate-$L$ regime (e.g., $L\geq 8$), which validates \emph{Proposition~\ref{TheoRD}}. As expected, the mismatch between the ``Theoretical'' and ``Empirical'' results diminishes as $L$ increases.

		\begin{figure}[!t]
			\centering
			\includegraphics[width=3.5in]{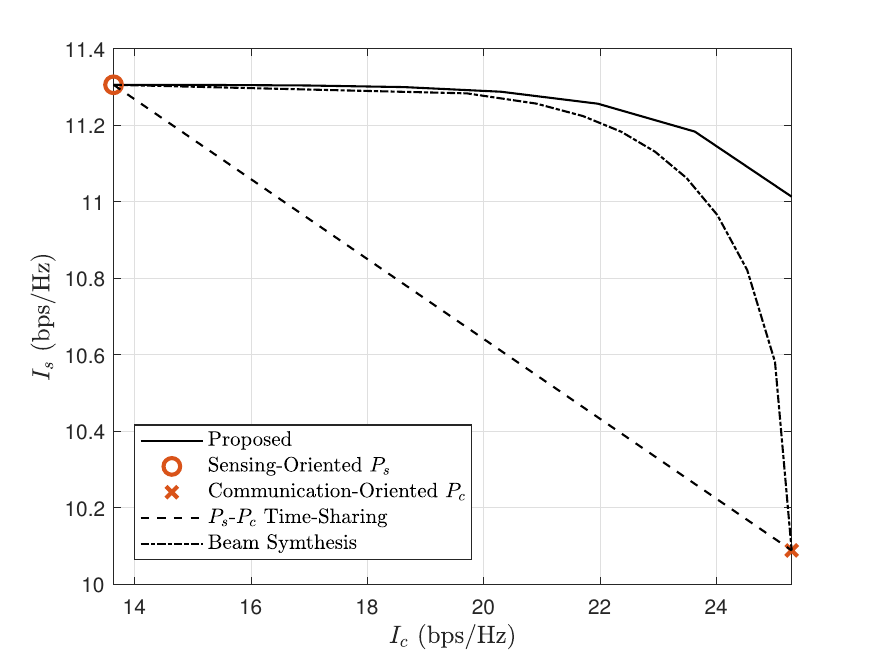}
			\caption{SMI versus the communication rate $I_c$.}
			\label{fig_comp}
		\end{figure}

	Fig. \ref{fig_comp} illustrates the trade-off between the SMI and the achievable communication rate by adjusting $R_0$. The curve labeled ``Proposed'' represents the pareto front achieved by the proposed precoding design, which optimally balances the dual objectives. The ``Sensing-Oriented $P_s$'' baseline represents the maximum achievable sensing performance obtained by directing the precoder entirely towards the sensing target and ignoring the communication user \cite{xie2020recursive}. The ``Communication-Oriented $P_c$'' baseline represents the precoder obtained by the waterfilling method, which maximizes the communication rate and ignores the sensing target \cite{kobayashi2006iterative}. The ``$P_s$-$P_c$ Time-Sharing'' denotes the inner bound obtained by	employing the time-sharing strategy between the two inner
	bounds $P_s$ and $P_c$ \cite{xiong2023fundamental}. The ``Beam Symthesis'' denotes the precoding design method proposed in \cite{xie2022perceptive}. 
	
	From Fig. \ref{fig_comp}, it is observed that the proposed method enables a flexible compromise between the dual functionalities of the ISAC system. As the requirement for the communication rate increases, the degrees of freedom available for sensing naturally decrease, leading to a reduction in SMI. 
Nevertheless, the proposed scheme consistently outperforms the benchmark candidates, achieving the highest SMI for any given communication rate.

	\section{Conclusions}
	In this paper, we investigated the performance limits of ISAC systems operating with practical hybrid waveforms. Unlike prior studies that isolated purely random or deterministic signals, our work analyzed a realistic signal model comprising both deterministic pilots and random data payloads. Then, we derived a tractable closed-form expression for the SMI, providing a precise information-theoretic metric that captures the impact of the hybrid signal. Based on this theoretical analysis, we developed a precoding design framework to maximize the SMI subject to strict transmit power and communication rate constraints, which is solved via an efficient ADMM-based algorithm. Simulation results demonstrated the accuracy of the derived asymptotic SMI expression and the superiority of the proposed precoding design.


	\end{document}